\documentstyle[sprocl]{article}

\input{psfig}

\def \ms {{\overline{\mbox{MS}}}}

\def\be{\begin{equation}}
\def\ee{\end{equation}}
\def\bea{\begin{eqnarray}}
\def\eea{\end{eqnarray}}

\begin{document}

\title{SMALL $X$ BEHAVIOUR OF PARTON DISTRIBUTIONS 
}

\author{A. V. KOTIKOV}

\address{Univ. Hamburg,
Institut f\"ur Theor.
Physik II\\
22761 Hamburg, Germany,~
E-mail: kotikov@mail.desy.de}

\author{G. PARENTE}

\address{Dep.
de F\'\i sica de Part\'\i culas, 
Univ.
de Santiago de Compostela\\
15706 Santiago de Compostela, Spain,~
E-mail: gonzalo@fpaxp1.usc.es}


\maketitle\abstracts{ 
We investigate  the  $Q^2$ evolution of parton distributions
at small $x$ values,
recently obtained in the case
of soft initial conditions.
The results are in excellent agreement with deep inelastic scattering
experimental data from HERA.}

\vskip -0.5cm
The measurements of the deep-inelastic scattering
structure function
$F_2$ in HERA
\cite{H1}
have permitted the access to
a very interesting kinematical range for testing the theoretical
ideas on the behavior of quarks and gluons carrying a very low fraction
of momentum of the proton, the so-called small $x$ region.
In this limit one expects that
non-perturbative effects may give essential contributions. However, the
resonable agreement between HERA data and the NLO approximation of
perturbative
QCD that has been observed for $Q^2 > 1 $GeV$^2$ (see the recent review
in \cite{CoDeRo}) indicates that
perturbative QCD could describe the
evolution of structure functions up to very low $Q^2$ values,
traditionally explained by soft processes.

Here we 
illustrate the results obtained recently in
\cite{Q2evo}.
These results are the extension to the NLO QCD approximation of previous 
LO studies \cite{Rujula}.
The main ingredients are:

{\bf 1.} Both, the gluon and quark singlet densities are
presented in terms of two components ($'+'$ and $'-'$)
which are obtained from the analytical $Q^2$
dependent expressions of the corresponding ($'+'$ and $'-'$)
parton distributions moments.

{\bf 2.} The $'-'$ component is constant
at small $x$, whereas the 
$'+'$ component grows at $Q^2 \geq Q^2_0$ as 
$$\sim \exp{\left(2\sqrt{\left[
a_+\ln \left(
\frac{a_s(Q^2_0)}{a_s(Q^2)} \right) -
\left( b_+ +  a_+ \frac{\beta_1}{\beta_0} \right)
\Bigl( a_s(Q^2_0) - a_s(Q^2) \Bigr) \right] 
\ln \left( \frac{1}{x}  \right)} \right)},
$$
where the LO term $a_+ = 12/\beta_0$ and the NLO one $b_+ = 412f/(27\beta_0)$. 
Here the coupling constant
$a_s=\alpha_s/(4\pi)$, $\beta_0$ and $\beta_1$ are the first two 
coefficients of QCD 
$\beta$-function and $f$ is the number of active flavors.

We have
analyzed $F_2$ HERA data at small $x$ from the H1 coll.\cite{H1}.
The initial scale of the parton distributions was fixed
into the fits to $Q^2_0$ = 1 $GeV^2$, although later it was released
to study the sensitivity of the fit to the variation of this parameter.
The analyzed data region was restricted to $x<0.01$ to remain within the
kinematical range where our results are
accurate. 

Fig. 1 shows $F_2$ calculated from the fit
with Q$^2$ $>$ 1 GeV$^2$.
Only the lower $Q^2$ bins are shown.
One can observe that the NLO result (dot-dashed line)
lies closer to the data
than the LO curve (dashed line).
The lack of agreement between data and lines observed
at the lowest $x$ and $Q^2$ bins suggests
that the initial flat behavior should occur at $Q^2$ lower
than 1 GeV$^2$.
In order to study this point we have done the
analysis considering $Q_0^2$ as a free parameter.
Comparing the results of the fits (see \cite{Q2evo})
one can notice
the better agreement with the experiment of the
NLO curve at fitted $Q^2_0=0.55 GeV^2$ (solid curve) which
is more apparent at the lowest kinematical bins.\\

\vskip -0.5cm

\begin{figure}[t]
\psfig{figure=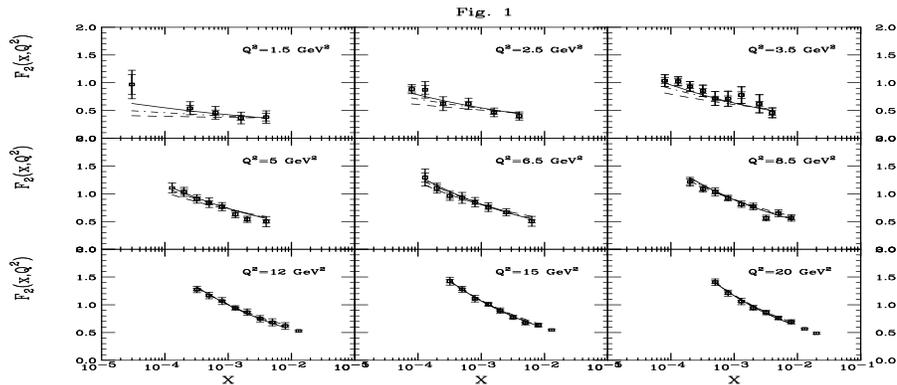,height=2.0in,width=4.7in}
\vskip -0.3cm
\caption{The structure function $F_2$ as a function of $x$ for different
$Q^2$ bins. The experimental points are from H1. 
The inner error 
bars are statistic while the outer bars represent statistic and systimatic 
errors added in quadrature. The dashen and dot-dashed curves are obtained 
from fits at LO and NLO respectively with fixed $Q^2_0=1$ GeV$^2$. The solid
line is from the fit at NLO giving $Q^2_0=0.55$ GeV$^2$.
 $\Lambda_{\ms}(f=4) = 250$ MeV is fixed.}
\vskip -0.5cm
\end{figure}

A.K. was supported by Alexander von Humboldt fellowship
and by DIS2000 Orgcommittee. 
G.P. was supported in part by Xunta de Galicia
(XUGA-20602B98) and CICYT (AEN96-1673).

{\bf References}

\end{document}